\documentclass[a4paper,11pt]{article}
\usepackage{pos,epsfig,bm}
\allowdisplaybreaks

\title{
\vspace*{-3.2cm}
\begin{minipage}{\textwidth}
{\normalfont\small LTH 1089, DESY 16-110, Nikhef 2016-030
\hspace{\fill} June 2016}\\
\end{minipage}\\[70pt]
Non-singlet coefficient functions for charged-current$\!\!$\\ 
deep-inelastic scattering to the third order in QCD}

\ShortTitle{Non-singlet coefficient functions for charged-current DIS}

\author*[a]{J. Davies}
\author[b]{S. Moch}
\author[c]{J.A.M. Vermaseren$\:\!$}
\author[a]{A. Vogt$\:\!$}

\affiliation[a]{Department of Mathematical Sciences, 
        University of Liverpool, Liverpool L69 3BX, UK}

\affiliation[b]{II. Institute for Theoretical Physics, 
        University of Hamburg, D-22761 Hamburg, Germany}

\affiliation[c]{Nikhef Theory Group, 
        Science Park 105, 1098 XG Amsterdam, The Netherlands \\[5mm]}

\emailAdd{Joshua.Davies@liv.ac.uk}
\emailAdd{Sven-Olaf.Moch@desy.de}
\emailAdd{t68@nikhef.nl}
\emailAdd{Andreas.Vogt@liv.ac.uk}

\abstract{
We have calculated the coefficient functions for the structure functions $\F2$,
$F_L$ and $\F3$ in $\nu\!-\!\nubar$ charged-current deep-inelastic scattering
(DIS) at the third order in the strong coupling $\ass\:\!$, thus completing 
the description of unpolarized inclusive $W^\pm$-exchange DIS to this order of 
massless perturbative QCD. In this brief note, our new results are presented 
in terms of compact approximate expressions that are sufficiently accurate for 
phenomenological analyses. For the benefit of such analyses we also collect,
in a unified notation, the corresponding lower-order contributions and the 
flavour non-singlet coefficient functions for $\nu\!+\!\bar{\nu}$ charged-%
current DIS. 
The behaviour of all six third-order coefficient functions at small Bjorken-$x$ 
is briefly discussed.
}
         
\FullConference{XXIV International Workshop on Deep-Inelastic Scattering and 
Related Subjects\\ 11-15 April 2016, DESY Hamburg, Germany}

\newcommand{\beq}{\begin{equation}}
\newcommand{\eeq}{\end{equation}}
\newcommand{\bea}{\begin{eqnarray}}
\newcommand{\eea}{\end{eqnarray}}
\newcommand{\nn}{\nonumber}

\newcommand{\lsim}{\raisebox{-0.07cm}{$\:\stackrel{<}{{\scriptstyle
 \sim}}\: $} }

\newcommand{\ar}{a_{\rm s}}
\newcommand{\als}{\alpha_{\rm s}^{}}
\newcommand{\ass}{\alpha_{\rm s}}

\newcommand{\MSb}{$\overline{\mbox{MS}}$}

\def\nubar{{\bar{\nu}}}
\def\F#1{{F_{\,#1}}}
\def\Qs{{Q^{\, 2}}}

\def\cf{{C_F}}
\def\nf{{n^{}_{\! f}}}
\def\nfs{{n^{\,2}_{\! f}}}

\def\z#1{{\zeta_{#1}^{}}}
\def\DD#1{{{\cal D}_{\:\!#1}}}
\def\DDk{{{\cal D}_{\:\!k}}}

\begin{document}
\maketitle

\section{Introduction}
 
\vspace{-1.5mm}
\noindent
With six structure functions, $\F2$, $F_L$ and  $\F3$ for $W^+$ and $W^-$ 
exchange \cite{BurasRev,PDG2014}, inclusive charged-current DIS is an important
source of information on the parton structure of nucleons and nuclei and on 
Standard Model parameters such as the strong coupling $\als$ and the weak 
mixing $\sin^2\theta_W^{}$~\cite{NuTeV2001}. 
Future facilities, e.g., the LHeC \cite{LHeC2012}, will be required to fully
realize its phenomenological potential. Charged-current structure functions
are, at very small values of the Bjorken variable $x$, also of interest for
the scattering of high-energy cosmic neutrinos, see Ref.~\cite{IKK2011} and 
references therein.

\vspace{0.5mm}
Here we address the corresponding coefficient functions (mass-factorized 
partonic cross sections) in massless perturbative QCD. 
These functions are relevant also beyond DIS, e.g., for Higgs production
via vector boson fusion \cite{VBF12,VBF16}.
For recent progress on heavy-quark effects see Ref.~\cite{BFdF2016}.

\vspace{-2mm}
\section{\boldmath First-order and $\nu+\bar{\nu}$ non-singlet coefficient 
functions}

\vspace{-1.5mm}
\noindent
The first-order coefficient functions for unpolarized inclusive DIS were
derived in the early days of QCD, see Refs.~\cite{BurasRev,FP82}.
The second- and third-order contributions for the $\nu+\bar{\nu}$ charged-%
current case have been calculated in Refs.~\cite{SanchezGuillen:1991iq,%
vanNeerven:1991nn,Zijlstra:1991qc,Zijlstra:1992kj,Moch:1999eb,MVV5,MVV6,MVV10}.
Here we collect the corresponding flavour non-singlet results in the \MSb\ 
scheme for the standard choice $\mu_F^{} = \mu_R^{} = Q$ of the 
renormalization and factorization scales. 
The additional contributions at other values of $\mu_F^{}$ and$\,$/or 
$\mu_R^{}$ are determined by these results and the corresponding splitting 
functions \cite{MVV3,MVV4}; see, e.g., Eqs.~(2.16) -- (2.18) in Ref.~\cite{NV2}.

\vspace{0.5mm}
We denote the non-singlet quark coefficient functions for the charged-current
structure functions $F_{2,\,3,\,L}^{\,\nu p \pm \bar \nu p}(x,\Qs)$ in 
neutrino-proton DIS by $C_{a,\pm}$, and write their perturbative expansion as

\vspace*{-2.5mm}
{\small
\beq
\label{Cexp}
  C_{a,\pm}(x,\Qs) \:=\: \sum_{\,n\,=\,0}^{} 
  \, \ar^{\,n} \, c_{a,\pm}^{\,(n)}(x)
  \quad \mbox{with} \quad \ar \:\equiv\: \als(\Qs)/(4\:\!\pi) \; .
\eeq
}

\vspace*{-1.5mm}
\noindent
In this notation the zeroth- and first-order coefficient functions
are given by
{\small
\bea
\label{c23Llo} 
  c_{2, \pm}^{\,(0)}(x) & = &
  c_{3, \pm}^{\,(0)}(x) \:=\: \delta(x_1) 
\; , \quad
  c_{L, \pm}^{\,(0)}(x) \:=\: 0 
\; , \quad
  c^{\,(1)}_{\,L,\pm}(x) \:=\: 4\, \* \cf \, \* x
\; , \\
\label{c23as1}
  c_{2, \pm}^{\,(1)}(x) & = &
  \cf \* \{ 4\, \* \DD1 - 3\, \* \DD0 - (9 + 4\, \* \z2)\, \* \delta(x_1)
  - 2\, \* (1+x) \* (L_1 - L_0)
  - 4\, \* x_1^{-1} \* L_0 + 6 + 4\, \* x \} 
\; , \quad
\\
  c_{3, \pm}^{\,(1)}(x) & = & c_{2, \pm}^{\,(1)}(x)
  - 2\, \* \cf\, \* (1+x)
\eea
}
with $\cf = (n_c^{\,2}-1)/(2\:\!n_c) = 4/3$ in QCD. Here and below we use the
abbreviations
{\small
\beq
\label{abbrev}
  x_1^{} \: = \: 1-x        \; ,\quad
  L_0 \: = \: \ln\, x       \; ,\quad
  L_1 \: = \: \ln\, x_1^{}  \; ,\quad
  \DDk \: = \: [\, x_1^{-1} L_1^{\:\!k\,}]_+ \; ,
\eeq
}
where $[a(x)]_+$ denotes +-distributions defined via $\,\int_{\:\!0}^1 dx \: 
[a(x)]_+ \, f(x) \:\equiv\: \int_{\:\!0}^1 dx \: a(x) \,\{ f(x) - f(1) \} $.
 
\vspace{0.1mm}
The coefficient functions $c_{a,+}^{\,(n)}$ and $c_{a,-}^{\,(n)}$ differ at 
$n>1$. The 2$^{\rm nd}$- and 3$^{\rm rd}$-order contributions to the former 
quantities read, in an approximate but sufficiently accurate form given in 
Refs.~\cite{MVV5,MVV6,MVV10},
{\small
\bea
\label{c2pls2}
  c_{2,+}^{\,(2)}(x) & \cong & 
%
       128/9\, \* \DD3 
     - 184/3\, \* \DD2 
     - 31.1052\, \* \DD1 
     + 188.641\, \* \DD0
     - 338.513\, \* \delta (x_1) 
     - 17.74\, \* L_1^3 
  \nn \\[-0.5mm] & & \mbox{} 
     + 72.24\, \* L_1^2 
     - 628.8\, \* L_1
     - 181.0 
     - 806.7\, \* x 
     + L_0 \* L_1 \* ( 37.75\, \* L_0 - 147.1\, \* L_1)
  \nn \\ & & \mbox{} 
     + 0.719\, \* xL_0^4 
     - 28.384\, \* L_0 
     - 20.70\, \* L_0^2 
     - 80/27\, \* L_0^3
  \\[0.5mm] & & \hspace{-8mm} \mbox{}
  + \,\nf\,\* \big\{ \,
       16/9\, \* \DD2 
     - 232/27\, \* \DD1 
     + 6.34888\, \* \DD0
     + 46.8531\, \* \delta (x_1) 
     - 1.500\,\* L_1^2 + 24.87\,\* L_1 
  \nn \\ & & \mbox{} 
     - 7.8109 
     - 17.82\, \* x 
     - 12.97\, \* x^2 
     + 8.113\, \* L_0 \* L_1
     - 0.185\, \* xL_0^3
     + 16/3\, \* L_0 + 20/9\, \* L_0^2 \,\big\}
\; , \;\; \nn \\
\label{cLpls2}
 c_{L,+}^{\,(2)}(x) &\cong& \mbox{}
%
     - 37.338 
     + 89.53\, \* x 
     + 33.82\, \* x^2
     + 128/9\, \* xL_1^2 
     - 46.50\, \* xL_1
     + xL_0 \* (32.90 + 18.41\, \* L_0)
 \nn \\[-0.5mm] & & \mbox{}
     - 84.094\, \* L_0 \* L_1
     - 128/9\, \* L_0 - 0.012\, \* \delta(x_1)
     \:+\: 16/27\, \* \nf\, \* 
     \big\{ 6\,\* xL_1 - 12\,\* xL_0 - 25\,\* x + 6 \big\} 
\; , \\
\label{c3pls2}
  c_{3,+}^{\,(2)}(x) & \cong &  
       128/9\, \* \DD3 
     - 184/3\, \* \DD2 
     - 31.1052\, \* \DD1 
     + 188.641\, \* \DD0
     - 338.572\, \* \delta (x_1^{}) 
     - 16.40\, \* L_1^3
  \nn \\[-0.5mm] & & \mbox{} 
     + 78.46\, \* L_1^2 - 470.6\, \* L_1
     - 149.75 
     - 693.2\, \* x 
     + 0.218\, \* xL_0^4
     + L_0 \* L_1 \* ( 33.62\, \* L_0 - 117.8\, \* L_1)
  \nn \\ & & \mbox{} 
     - 49.30\, \* L_0 
     - 94/3\, \* L_0^2 
     - 104/27\, \* L_0^3
  \\[0.5mm] & & \hspace{-8mm} \mbox{} 
  + \,\nf \, \* \big\{ \,
       16/9\, \* \DD2 
     - 232/27\, \* \DD1 
     + 6.34888\, \* \DD0
     + 46.8464\, \* \delta (x_1^{}) 
     + 0.066\, \* L_1^3 
     - 0.663\, \* L_1^2 
  \nn \\ & & \mbox{} 
     + 24.86\, \* L_1 
     - 5.738 
     - 5.845\, \* x 
     - 10.235\, \* x^2 
     - 0.190\, \* xL_0^3
     + 4.265\, \* L_0 \* L_1 
     + 20/9\, \* L_0 \* ( 4 + L_0 ) \big\} 
\nn
 \eea
 }
  
 \vspace*{-8mm}
 \noindent
 and
 {\small
 \bea
\label{c2ns3}
  c_{2,+}^{\,(3)}(x) & \cong &  \;
       512/27\, \* \DD5 
     - 5440/27\, \* \DD4 
     + 501.099\, \* \DD3 
     + 1171.54\, \* \DD2 
     - 7328.45\, \* \DD1 
     + 4442.76\, \* \DD0
  \nn \\[-0.5mm] & & \mbox{} 
     - 9170.38\, \*\delta (x_1) 
     - 512/27\, \* L_1^5 
     + 704/3\, \* L_1^4 
     - 3368\, \* L_1^3 
     - 2978\, \* L_1^2 
     + 18832\, \* L_1
     - 4926 
  \nn \\ & & \mbox{} 
     + 7725\, \* x 
     + 57256\, \* x^2 
     + 12898\, \* x^3
     - 56000\, \*x_1 \*L_1^2 
     - L_0 \* L_1 \* (6158 + 1836\, \* L_0) 
     + 4.719\, \* xL_0^5
  \nn \\ & & \mbox{} 
     - 775.8\, \* L_0 
     - 899.6\, \* L_0^2 
     - 309.1\, \* L_0^3 
     - 2932/81\, \* L_0^4 
     - 32/27\, \* L_0^5 
  \nn \\[0.5mm] & & \hspace{-8mm} \mbox{} 
  + \,\nf\, \* \big\{ \;
       640/81\, \* \DD4 
     - 6592/81\, \* \DD3 
     + 220.573\, \* \DD2
     + 294.906\, \* \DD1
     - 729.359\, \* \DD0 
  \nn \\ & & \mbox{} 
     + 2574.687\, \* \delta (x_1) 
     - 640/81\, \* L_1^4 
     + 153.5\, \* L_1^3 
     - 828.7\, \* L_1^2 
     - 501.1\, \* L_1
     + 831.6 
     - 6752\, \* x 
  \nn \\ & & \mbox{} 
     - 2778\, \* x^2 
     + 171.0\, \* x_1 \* L_1^4
     + L_0 \* L_1\, \* (4365 + 716.2\, \* L_0 - 5983\, \* L_1) 
     + 4.102\, \* xL_0^4
     + 275.6\, \* L_0 
  \nn \\ & & \mbox{} 
     + 187.3\, \* L_0^2 
     + 12224/243\, \* L_0^3 
     + 728/243\, \* L_0^4 \, \big\}
  \\[0.5mm] & & \hspace{-8mm} \mbox{} 
  + \,\nfs\, \* \big\{ \;
       64/81\, \* \DD3 
     - 464/81\, \* \DD2 
     + 7.67505\, \* \DD1 
     + 1.00830\, \* \DD0
     - 103.2366\, \* \delta (x_1)
     - 64/81\, \* L_1^3 
  \nn \\ & & \mbox{} 
     + 18.21\, \* L_1^2 
     - 19.09\, \* L_1
     + 129.2\, \* x 
     + 102.5\, \* x^2 
     + L_0 \* L_1\, \* (- 96.07 - 12.46\, \* L_0 + 85.88\, \* L_1)
  \nn \\ & & \mbox{} 
     - 8.042\, \* L_0 
     - 1984/243\, \* L_0^2 
     - 368/243\, \* L_0^3 \, \big\}
%
\; , \nn \\[2mm]
\label{cLpls3}
 c^{\,(3)}_{L,+}(x) & \cong & \;
      512/27\, \* L_1^4 
    - 177.40\, \* L_1^3 
    + 650.6\, \* L_1^2 
    - 2729\, \* L_1
    - 2220.5 - 7884\, \* x 
    + 4168\, \* x^2
  \nn  \\[-0.5mm] & & \mbox{}
    - (844.7\, \* L_0 + 517.3\, \* L_1) \* L_0 \* L_1
    + (195.6\, \* L_1 - 125.3)\, \* x_1 \* L_1^3
    + 208.3\, \* xL_0^3 
    - 1355.7\, \* L_0
  \nn  \\ & & \mbox{}
    - 7456/27\, \* L_0^2 
    - 1280/81\, \* L_0^3
    + 0.113\, \* \delta(x_1)
  \nn \\[0.5mm] & & \hspace{-8mm} \mbox{} 
  + \,\nf\, \* \big\{
      1024/81\, \* L_1^3 
    - 112.35\, \* L_1^2 
    + 344.1\, \* L_1 
    + 408.4 
    - 9.345\, \* x
    - 919.3\, \* x^2
  \nn  \\ & & \mbox{}
    + (239.7 + 20.63\, \* L_1)\, \* x_1 \* L_1^2
    + (887.3 + 294.5\, \* L_0 - 59.14\, \* L_1) \* L_0 \* L_1
    - 1792/81\, \* xL_0^3
   \nn  \\ & & \mbox{}
    + 200.73\, \* L_0 
    + 64/3\, \* L_0^2
    + 0.006\,\* \delta(x_1) \big\}
  \\[0.5mm] & & \hspace{-8mm} \mbox{} 
  + \,\nfs\, \* \big\{
      3\, \* xL_1^2 + ( 6 - 25 \* x) \* L_1 - 19 
    + (317/6 - 12\, \* \zeta_2)\,\* x 
    - 6\, \* xL_0 \* L_1
    + 6\*x\, \* \mbox{Li}_2(x)
    + 9\, \* xL_0^2
  \nn  \\ & & \mbox{}
    - (6 - 50 \* x) \* L_0 \big\} \, \* 64/81
  \; , \nn
\\[2mm]
\label{c3pls3}
  c_{3,+}^{\,(3)}(x) & \cong & \; 
%
       512/27\, \* \DD5 
     - 5440/27\, \* \DD4 
     + 501.099\, \* \DD3
     + 1171.54\, \* \DD2 
     - 7328.45\, \* \DD1 
     + 4442.76\, \* \DD0
  \nn \\[-0.5mm] && \mbox{} 
     - 9172.68\, \* \delta (x_1^{})
     - 512/27\, \* L_1^5 
     + 8896/27\, \* L_1^4 
     - 1396\, \* L_1^3
     + 3990\, \* L_1^2 
     + 14363\, \* L_1
  \nn \\ & & \mbox{} 
     - 1853 
     - 5709\, \* x 
     + x\,\* x_1^{} \* (5600 - 1432\, \* x)
     - L_0 \* L_1 \* (4007 + 1312\, \* L_0) 
     - 0.463\, \* xL_0^6 
  \nn \\ & & \mbox{} 
     - 293.3\, \* L_0 
     - 1488\, \* L_0^2 
     - 496.95\, \* L_0^3 
     - 4036/81\, \* L_0^4 
     - 536/405\, \* L_0^5
  \nn \\[0.5mm] & & \hspace{-8mm} \mbox{} 
  + \,\nf\, \* \big\{
       640/81\, \* \DD4 
     - 6592/81\, \* \DD3 
     + 220.573\, \* \DD2
     + 294.906\, \* \DD1 
     - 729.359\, \* \DD0 
  \nn \\ & & \mbox{} 
     + 2575.46\, \*\delta (x_1^{}) 
     - 640/81\, \* L_1^4 + 32576/243\, \* L_1^3 
     - 660.7\, \* L_1^2 + 959.1\, \* L_1 
     + 31.95\, \* x_1^{} \* L_1^4
  \nn \\ & & \mbox{} 
     + 516.1
     - 465.2\, \* x 
     + x\,\* x_1^{} \* (635.3 + 310.4\, \* x)
     + L_0 \* L_1\, \* (1496 + 270.1\,\* L_0 - 1191\,\*  L_1)
  \nn \\ & & \mbox{} 
     - 1.200\, \* xL_0^4 
     + 366.9\,\*  L_0
     + 305.32\, \* L_0^2 
     + 48512/729\,\*  L_0^3 
     + 304/81\, \* L_0^4 \, \big\}
  \\[0.5mm] & & \hspace{-8mm} \mbox{} 
  + \,\nfs\, \*  \big\{
       64/81\, \* \DD3 
     - 464/81\, \* \DD2 
     + 7.67505\, \* \DD1 
     + 1.00830\, \* \DD0
     - 103.2602\, \* \delta (x_1^{}) 
     - 64/81\, \* L_1^3
  \nn \\ & & \mbox{} 
     + 992/81\, \* L_1^2 
     - 49.65\,\* L_1 
     + 11.32 - x\,\* x_1^{}\* (44.52 + 11.05\, \* x)
     + 51.94\,\*  x
     + 0.0647\,\* xL_0^4
  \nn \\ & & \mbox{} 
     - L_0 \* L_1\, \* ( 39.99 + 5.103\, \* L_0 - 16.30\, \* L_1)
     - 16.00\, \* L_0 
     - 2848/243\, \* L_0^2 
     - 368/243\, \* L_0^3 \, \big\}
  \nn \\[0.5mm] & & \hspace{-8mm} \mbox{} 
  + fl_{02} \,\* \nf\, \* \big\{
       2.147\, \* L_1^2 
     - 24.57\, \* L_1 
     + 48.79 
     - x_1^{} \* (242.4 - 150.7\, \* x)
     - L_0 \* L_1\, \* (81.70 + 9.412\, \* L_1)
  \nn \\ & & \mbox{} 
     + xL_0 \, \* (218.1 + 82.27\,\* L_0^2)
     - 477.0\, \* L_0 
     - 113.4\, \* L_0^2  
     + 17.26\, \* L_0^3
     - 16/27\, \* L_0^5
     \,\big\} \, \* x_1^{}
\; . \nn
\eea
}
 
\vspace*{-1cm}
\section{\boldmath $\nu-\bar{\nu}$ non-singlet coefficient functions}

\vspace*{-1mm}
\noindent
The differences between corresponding $\nu+\bar{\nu}$ and $\nu-\bar{\nu}$
coefficient functions are, as conjectured in Ref.~\cite{BKM04}, suppressed 
at large $x$ by two powers of $1\!-\!x$. Hence it is convenient to present the 
coefficient functions for $\nu-\bar{\nu}$ charged-current DIS in terms of 
differences which we define as
\bea
  \label{c23Ldff}
  \delta\:\! C_{2,L} \; \equiv \; C_{2,L}^{\,\nu p + {\bar \nu} p}
    - C_{2,L}^{\,\nu p - {\bar \nu} p} \:\: , \qquad
  \delta\:\! C_3 \; \equiv \;  C_3^{\,\nu p - {\bar \nu} p}
    - C_3^{\,\nu p + {\bar \nu} p} \; .
\eea
The flavour class $fl_{02\,}$, see Fig.~1 of Ref.~\cite{MVV6}, does not 
contribute to the flavour asymmetries probed in the $\nu-\bar{\nu}$ 
combinations, hence it is understood that the corresponding part of
Eq.~(\ref{c3pls3}) is removed before the difference for $F_3$ is formed.
The $\delta\:\! C_{a}$ can be perturbatively expanded as
\bea
\label{cf-exp}
  \delta\:\! C_a \; = \;
  \sum_{n=2} \: \ar^{\, n}\: \delta\:\! c_{a}^{(n)} \; ,
\eea
where $\ar$ is defined in Eq.~(\ref{Cexp}) above. The second-order
results were already given in Ref.~\cite{MVRogal07} in exact and
parametrized form. The later results, written in terms of the 
abbreviations (\ref{abbrev}), read 
\bea
\label{dc2ns2p}
 \delta c_{2}^{(2)}(x) & \cong &
%
 \{ 
   - 9.1587 
   - 57.70\,\* x 
   + 72.29\, \* x^2 
   - 5.689\, \* x^3
   - xL_0\, \* ( \, 68.804 + 24.40\, \* L_0 
 \nn \\ & & \mbox{} \;
     + 2.958\, \* L_0^2 \, )
   + 0.249\, \* L_0 
   + 8/9\: \* L_0^2\,\* (2 + L_0) \, 
 \} \, \* x_1
 \; , \\
\label{dcLns2p}
 \delta c_{L}^{(2)}(x) & \cong &
%
  \{ \,
     10.663
   - 5.248\, \* x
   - 7.500\, \* x^2
   + 0.823\, \* x^3
   + xL_0\, \* ( \, 11.10 + 2.225\, \* L_0
 \nn \\ & & \mbox{} \;
   - 0.128\, \* L_0^2 \: )
   + 64/9\: \* L_0 \,
  \} \, \* x_1^2
 \; , \\
\label{dc3ns2p}
 \delta c_{3}^{(2)}(x) & \cong &
%
  \{ 
   - 29.65
   + 116.05\, \* x 
   - 71.74\, \* x^2 
   - 16.18\, \* x^3
   + xL_0\,\*  ( \, 14.60 + 69.90\, \* x
 \nn \\ & & \mbox{}
   - 0.378\, \* L_0^2 \, )
   - 8.560\, \* L_0 
   + 8/9\: \* L_0^2\, \* (4 + L_0) \, 
 \} \: \* x_1
 \; .  
\eea
 
The corresponding third-order corrections are the new results of the present
contribution. They supersede the previous approximate expressions in Eqs.~(3.7)
-- (3.9) of Ref.~\cite{MVRogal07}, which were based on the lowest five even-
integer and odd-integer Mellin moments of $C_{3,-}$ and $C_{a,-}$, $a=2,\,L$,
respectively, computed in Ref.~\cite{MRogal07}. 
Our new exact results can parametrized as
\bea
\label{dc2ns3p}
 \delta c_{2}^{\,(3)} & \cong & 
%
 \big\{ \, 
     273.59 
   - 44.95\, \* x 
   - 73.56\, \* x^2 
   + 40.68\, \* x^3 
   + 0.1356\, \* L_0^5
   + 8.483\, \* L_0^4 
   + 55.90\, \* L_0^3
\nn\\ & & \hspace{2mm}
   + 120.67\, \* L_0^2 
   + 388.0\, \* L_0 
   - 329.8\, \* L_0\* L_1 
   - x\* L_0 \* (316.2 + 71.63\, \* L_0) 
   + 46.30\, \* L_1
\nn\\ & & \hspace{2mm}
   + 5.447\, \* L_1^2
 \big\} \, \* x_1 
   - 0.0008\,\* \delta(x_1)
\\& & \hspace{-9mm} \mbox{}
 + \,\nf\, \* \big\{ \big(
   - 19.093\, 
   + 12.97\, \* x 
   + 36.44\, \* x^2 
   - 29.256\, \* x^3 
   - 0.76\, \* L_0^4 
   - 5.317\, \* L_0^3 
   - 19.82\, \* L_0^2
\nn\\ & & \hspace{2mm}
   - 38.958\, \* L_0 
   - 13.395\, \* L_0 \* L_1 
   + xL_0 \* (14.44 + 17.74\, \* L_0) 
   + 1.395\, \* L_1
 \big) \, \* x_1 
   + 0.0001\, \* \delta(x_1) 
 \big\}
\nn \; , \\[1mm]
\label{dcLns3p}
 \delta c_{L}^{\,(3)} & \cong &
 \big\{
   - 620.53 
   - 394.5\, \* x 
   + 1609\, \* x^2 
   - 596.2\, \* x^3 
   + 0.217\, \* L_0^3 
   + 62.18\, \* L_0^2 
   + 208.47\, \* L_0 
\nn\\ & & \hspace{2mm} \mbox{} 
   - 482.5\, \* L_0 \* L_1 
   - xL_0 \* (1751 - 197.5\, \* L_0) 
   + 105.5\, \* L_1 
   + 0.442\, \* L_1^2 
 \big\} \, \* x_1^2
\\ & & \hspace{-9mm} \mbox{} 
 + \,\nf\, \* \big\{
   - 6.500 
   - 12.435\, \* x 
   + 23.66\, \* x^2 
   + 0.914\, \* x^3
   + 0.015\, \* L_0^3 
   - 6.627\, \* L_0^2  
   - 31.91\, \* L_0
\nn\\ & & \hspace{2mm} \mbox{}
   - xL_0 \* (5.711 + 28.635\, \* L_0)
 \big\}\, \* x_1^2
\nn \; , \\[1mm]
\label{dc3ns3p}
 \delta c_{3}^{\,(3)} & \cong & 
%
 \big\{ 
   - 553.5 
   + 1412.5\, \* x 
   - 990.3\, \* x^2 
   + 361.1\, \* x^3 
   + 0.1458\, \* L_0^5 
   + 9.688\, \* L_0^4 
   + 90.62\, \* L_0^3
\nn\\ & & \hspace{2mm} \mbox{}
   + 83.684\, \* L_0^2 
   - 602.32\, \* L_0 
   - 382.5\, \* L_0 \* L_1 
   - x\* L_0 \* (2.805 + 325.92\, \* L_0) 
   + 133.5\, \* L_1
\nn\\ & & \hspace{2mm} \mbox{}
   + 10.135\, \* L_1^2
 \big\} \, \* x_1 
   - 0.0029\, \* \delta(x_1)
\\ & & \hspace{-9mm} \mbox{}
 + \,\nf\, \* \big\{ \big( 
   - 16.777 
   + 77.78\, \* x 
   - 24.81\, \* x^2 
   - 28.89\, \* x^3 
   - 0.7714\, \* L_0^4 
   - 7.701\, \* L_0^3 
   - 21.522\, \* L_0^2
\nn\\ & & \hspace{2mm}
   - 7.897\, \* L_0 
   - 16.17\, \* L_0 \* L_1 
   + xL_0 \* (43.21 + 67.04\, \* L_0) 
   + 1.519 \* L_1 
 \big) \, \* x_1
   + 0.00006\, \* \delta(x_1) 
 \big\}
\nn \; .
\eea

\vspace*{-1mm}
\section{Discussion}

\vspace*{-1.5mm}
\noindent
With the exception of the $\nf$ part of Eq.~(\ref{cLpls2}) and the $\nfs$ part
of Eq.~(\ref{cLpls3}) which are exact, the second- and third-order expressions 
in sections 2 and 3 have been obtained by fitting the coefficients not written
as fractions in the non-distribution parts to the exact coefficient functions 
at $x \geq 10^{\,-6}$. Where useful, the coefficients of $\delta(1\!-\!x)$ have 
been adjusted (even from zero) to fine-tune the accuracy of Mellin moments and 
convolutions. 
The resulting accuracy of Eqs.~(\ref{c2pls2}) -- (\ref{c3pls3}) and 
(\ref{dc2ns2p}) -- (\ref{dc3ns3p}) and their convolutions with typical 
quark distributions of hadrons is 0.1\% or better except where the functions 
are very small. Towards smaller $x$ the accuracy deteriorates, but the results 
are still accurate to about 1\% and 3\% at $x = 10^{\,-8}$ and $x = 10^{\,-10}$,
respectively.

\vspace{1mm}
{\sc Fortran} subroutines of these functions can be obtained from the preprint 
server$\,$ {\sc arXiv.org} by downloading the source of this note. They are 
also available from the authors upon request.

\vspace{1mm}
Analogous parametrizations for the pure-singlet quark and gluon coefficient
functions for $\F2$ and $F_L$ have been given in Ref.~\cite{MVV5} and 
section~4 of Ref.~\cite{MVV6}. The partly very lengthy exact expression 
corresponding to Eqs.~(\ref{c2pls2}) -- (\ref{c3pls3}) and (\ref{dc2ns2p}) -- 
(\ref{dc3ns2p}) can be found in Ref.~\cite{MVV6} -- where the $fl_{11}$
contribution has to be disregarded for the present charged-current case --
and Refs.~\cite{MVV10,MVRogal07}; those for Eqs.~(\ref{dc2ns3p}) -- 
(\ref{dc3ns3p}) can be found in Ref.~\cite{JDthesis}. Only the latter 
expressions allow for the analytical calculation of all integer Mellin moments 
of the coefficient functions. 

\vspace{1mm}
The second moments of $\delta c_{2}^{\,(3)}$ and $\delta c_{L}^{\,(3)}$ are of
particular relevance, since they enter the QCD corrections to the 
Paschos-Wolfenstein relation \cite{PaschosW} for the determination of 
$\sin^2 \theta_W^{}$ from charged-current DIS \cite{NuTeV2001}. 
The truncated numerical values of these moments for $\nf$ light flavours are
%
\beq
\label{moment2}
  \delta \:\! c_{2}^{(3)}(N=2) \;=\; - 20.4001 + 0.72202\, \* \nf 
\; , \quad 
  \delta \:\! c_{L}^{(3)}(N=2) \;=\; - 24.7755 + 0.80134\, \* \nf
\; .
\eeq
Within their error estimates, the previous approximate results for these 
moments \cite{MVRogal07} agree with Eq.~(\ref{moment2}). The corresponding 
analytical expressions can also be found in Ref.~\cite{JDthesis}.
 
\begin{figure}[hbt]
\vspace*{-2mm}
\centerline{\hspace*{-2mm}\epsfig{file=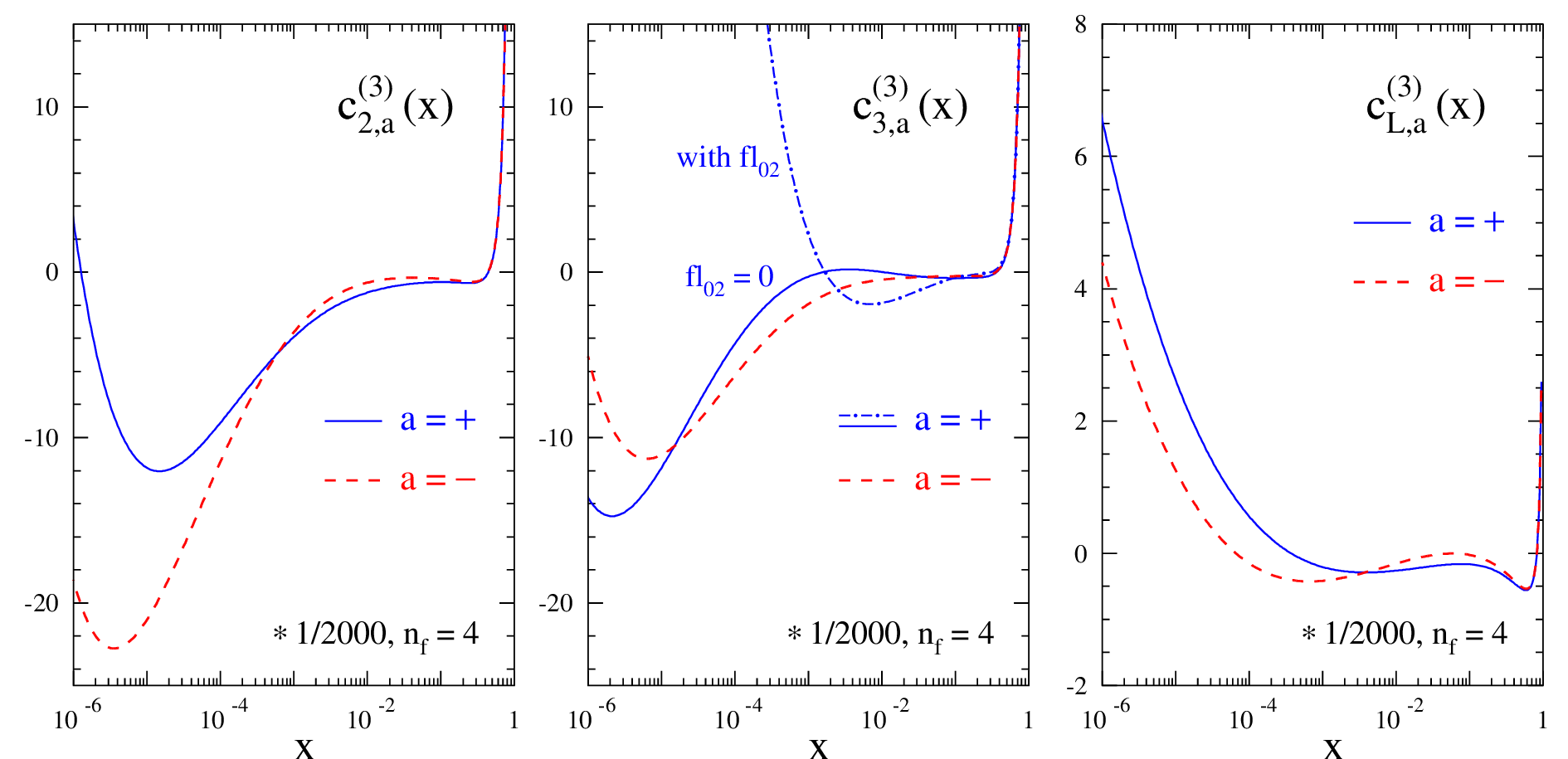,width=15.5cm,angle=0}}
\vspace{-2mm}
\caption{\label{Fig4}
The third-order non-singlet coefficient functions for $\F2$, $F_L$ and $\F3$ 
in $\nu p \!+\! \bar{\nu} p$ ($\,a = +$) and $\nu p \!-\! \bar{\nu} p$ 
($\,a = -$) charged-current DIS for four light flavours. 
The function $c_{3,+}^{\,(3)}(x)$ in Eq.~(\protect\ref{c3pls3}) is shown with 
and without the $fl_{02}$ contribution. 
The factor $1/2000$ approximately converts the curves to 
an expansion in $\ass$.
}
\vspace{-3mm}
\end{figure}

The $a_{\rm s}^{\,3}$ coefficients $c_{a,\pm}^{\,(3)}$ in Eq.~(\ref{Cexp}) 
are illustrated in Fig.~1 over a wide range in $x$. All~six functions exhibit
a sharp small-$x$ rise, but only at $x\!<\! 10^{-5}$ for $\F2$ and $\F3$ and at
$x \lsim 10^{-4}$ for~$F_{L\,}$. 
With the~exception of the flavour structure $fl_{02}$ that occurs at three 
loops for the first time but dom\-inates $c_{a,+}^{\,(3)}$ at small $x$, at 
least four of five $\ln^{\,k\!} x$ small-$x$ terms are required for a good 
approximation for $c_{2,\pm}^{\,(3)}$ and $c_{3,\pm}^{\,(3)}$, and all three 
such terms for $c_{L,\pm}^{\,(3)}$, even below the $x$-range shown in the 
figure.
Further discussions and illustrations of these coefficient functions can be
found in Ref.~\cite{JDthesis}.

\vspace{-2mm}
\section*{Acknowledgements}

\vspace*{-2mm}
\noindent
This work has been supported by the UK {\it Science \& Technology Facilities 
Council}$\,$ (STFC) grants ST/L000431/1 and ST/K502145/1, 
by the {\it Deutsche Forschungsgemeinschaft} (DFG) via contract MO 1801/1-1,
and by the {\it European Research Council}$\,$ (ERC) Advanced Grant 320651, 
{\sc Hepgame}.


{\small 
\vspace*{-2mm}
\begin{thebibliography}{99}

\setlength{\itemsep}{1.4mm}

\bibitem{BurasRev}
A.J. Buras, 
  Rev.\ Mod.\ Phys.\ 52 (1980) 199, and references therein

\bibitem{PDG2014}
  K.A.~Olive et al., Particle Data Group,
  Chin.\ Phys.\ C38 (2014) 090001

\bibitem{NuTeV2001}
G.~Zeller {\it et al.}, NuTeV,
  Phys.\ Rev.\ Lett.\ 88 (2002) 091802
  E: {\it ibid.} 90 (2003) 239902, hep-ex/0110059$\!\!\!\!$

\bibitem{LHeC2012}
J.L.~Abelleira Fernandez {\it et al.}, LHeC Study Group,
  J.\ Phys.\ G39 (2012) 075001, arXiv:1206.2913

\bibitem{IKK2011}
A.Y.~Illarionov, B.A.~Kniehl and A.V.~Kotikov,
  Phys.\ Rev.\ Lett.\ 106 (2011) 231802, arXiv:1105.2829 

\bibitem{VBF12}
P.~Bolzoni, F.~Maltoni, S.~Moch and M.~Zaro,
  Phys.\ Rev.\ D85 (2012) 035002, arXiv:1109.3717

\bibitem{VBF16}
F.A.~Dreyer and A.~Karlberg, arXiv:1606.00840

\bibitem{BFdF2016}
  J.~Bl\"umlein, G.~Falcioni and A.~De Freitas,
  arXiv:1605.05541

\bibitem{FP82}
W. Furmanski and R. Petronzio,
  Z. Phys.\ C11 (1982) 293, and references therein

\bibitem{SanchezGuillen:1991iq}
J. Sanchez Guillen et~al.,
  Nucl.\ Phys.\ B353 (1991) 337

\bibitem{vanNeerven:1991nn}
W.L. van Neerven and E.B. Zijlstra,
  Phys.\ Lett.\ B272 (1991) 127

\bibitem{Zijlstra:1991qc}
E.B. Zijlstra and W.L. van Neerven,
  Phys.\ Lett.\ B273 (1991) 476

\bibitem{Zijlstra:1992kj}
E.B. Zijlstra and W.L. van Neerven,
  Phys.\ Lett.\ B297 (1992) 377

\bibitem{Moch:1999eb}
S. Moch and J.A.M. Vermaseren,
  Nucl.\ Phys.\ B573 (2000) 853, hep-ph/9912355

\bibitem{MVV5}
  S.~Moch, J.A.M.~Vermaseren and A.~Vogt,
  Phys.\ Lett.\ B606 (2005) 123, hep-ph/0411112

\bibitem{MVV6}
J.A.M.~Vermaseren, A.~Vogt and S.~Moch,
  Nucl.\ Phys.\ B724 (2005) 3, hep-ph/0504242

\bibitem{MVV10}
S. Moch, J.A.M. Vermaseren and A.~Vogt,
  Nucl.\ Phys.\ B813 (2009) 220, arXiv:0812.4168 

\bibitem{MVV3}
S. Moch, J.A.M. Vermaseren and A. Vogt,
  Nucl.\ Phys.\ B688 (2004) 101, hep-ph/0403192

\bibitem{MVV4}
A. Vogt, S. Moch and J.A.M. Vermaseren,
  Nucl.\ Phys.\ B691 (2004) 129, hep-ph/0404111

\bibitem{NV2}
W.L. van Neerven and A. Vogt,
  Nucl.\ Phys.\ B588 (2000) 345, hep-ph/0006154

\bibitem{MVRogal07}
S.~Moch, M.~Rogal and A.~Vogt,
  Nucl.\ Phys.\ B790 (2008) 317, arXiv:0708.3731

\bibitem{MRogal07}
S.~Moch and M.~Rogal,
  Nucl.\ Phys.\ B782 (2007) 51, arXiv:0704.1740

\bibitem{BKM04}
D.J. Broadhurst, A.L. Kataev and C.J. Maxwell,
  Phys.\ Lett.\ B590 (2004) 76, hep-ph/0403037

\bibitem{PaschosW}
E.A. Paschos and L. Wolfenstein,
  Phys.\ Rev.\ D7 (1973) 91

\bibitem{JDthesis}
  J. Davies, 
  PhD thesis, University of Liverpool (2016), \\
  {\tt https://livrepository.liverpool.ac.uk/3003745/}

\end{thebibliography}
}
\end{document}